\documentclass{article}
\usepackage{spconf,amsmath,graphicx, hyperref,xcolor,colortbl,subfig, comment}
\usepackage{url}
\usepackage{flushend}
\usepackage[stable]{footmisc}

\hypersetup{
    colorlinks=true,
    citecolor=blue,
    linkcolor=blue,
    filecolor=magenta,      
    urlcolor=blue,
}

\title{MusicNet: Compact Convolutional Neural Network for Real-time Background Music Detection}
\name{
    Chandan K.A. Reddy,
    Vishak Gopal,
    Harishchandra Dubey,
    Sergiy Matusevych,
    Ross Cutler, 
    Robert Aichner
}

\address{
    Microsoft Corporation, Redmond, WA, USA \\
    chandan.ka@outlook.com, \{vigopal,hadubey,sergiym,rcutler,raichner\}@microsoft.com}

\begin{document}
%
\maketitle
%
%
\begin{abstract}
With the recent growth of remote work, online meetings often encounter challenging audio
contexts such as background noise, music, and echo. Accurate real-time detection of music events can
help to improve the user experience. In this paper, we present MusicNet, a compact neural model for detecting background music in the real-time communications pipeline. In video meetings, music frequently co-occurs with speech and background noises, making the accurate classification quite challenging. We propose a compact convolutional neural network core preceded by an in-model featurization layer. MusicNet takes 9 seconds of raw audio as input and does not require any model-specific featurization in the product stack. We train our model on the balanced subset of the Audio Set~\cite{gemmeke2017audio} data and validate it on 1000 crowd-sourced real test clips. Finally, we compare MusicNet performance with 20 state-of-the-art models. MusicNet has a true positive rate (TPR) of 81.3\% at a 0.1\% false positive rate (FPR), which is significantly better than state-of-the-art models included in our study. MusicNet is also 10x smaller and has 4x faster inference than the best performing models we benchmarked.
\end{abstract}
\vspace{-4mm}
\section{Introduction}
\label{sec:intro}
\vspace{-2mm}
Audio event detection is not a new challenge for machine perception. The term Audio Event Detection refers to the recognition of sound events in a variety of scenarios. Recently, researchers have proposed
several methods for audio event detection in different contexts, e.g., using weak labels, targeting
specific audio events, or in challenging signal-to-noise ratio conditions
\cite{papadimitriou2020audio, deshmukh2020multi}.
Detection of specific audio events can also be of interest. Tasks such as speech or voice activity
detection for recognizing the presence of human speech \cite{sohn1999statistical,
zhang2015boosting}, or music activity detection \cite{seyerlehner2007automatic,
izumitani2008background, de2019exploring, jia2020hierarchical}, have been extensively studied. This
paper proposes a solution for low-complexity real-time music detection in the presence of speech and
background noises.

Past work in the field typically involved rule-based systems, used relatively small datasets and
a reduced sets of labels, and assumed simpler event-specific settings for event detection and
classification \cite{stowell2015detection, temko2006clear}.

Google Audio Set~\cite{gemmeke2017audio} is a collection of labels associated with 10-second
fragments of YouTube videos. Each video fragment is associated with one or more labels from the
ontology of 632 sound categories, and the classification is weakly supervised. Audio Set consists of
more than two million unique clips, making it the largest publicly available corpus of labeled
sounds. The public release of the Audio Set data encouraged the development of deep neural networks for
audio event detection and classification~\cite{8461392, Kong_2019}.

We use Audio Set to train a binary classifier for robust music detection. We require a high true
positive rate at a very low FPR in the presence of speech, singing, background
noise, and in other challenging audio contexts. Specifically we require a FPR $\leq$ 0.1\% so that users don't receive a false positive more often that once every 3 hours of continuous evaluation in real-time audio calls, which we've determined sufficient for our application in a mainstream audio communication system. In addition, we require a TPR $\geq$ 80\% to make music detection useful. Our application also requires fast real-time
inference on edge devices, a small model size, and low deployment and maintenance costs of the model. Our
model contains in-model featurization\footnote{Featurization is the preprocessing of raw audio into features for the DNN classifier} where a 1-D convolutional layer learns to produce spectral
features from raw audio.

We validate the model on a realistic test set consisting of a blind test set from the 3rd Deep
Noise Suppression Challenge~\cite{ch2021interspeech} and a collection of real-world music clips
created via crowd-sourcing.

The main contributions of MusicNet are (1) It provides a best-in-class TPR evaluated at a low FPR needed for real-world usage, (2) We provide the most extensive evaluation of state-of-the-art music detectors we are aware of, (3) MusicNet has the lowest complexity and smallest model size of any high performing music detectors we've evaluated, (4) MusicNet includes in-model featurization for easy integration and maintenance into applications; no source code changes are needed to update model if the input features are changed. 

\vspace{-5mm}
\section{Related works}
\vspace{-3mm}
The Music Information Retrieval Evaluation eX-
change (MIREX) 2018 challenge~\cite{mirex_challenge} fueled the interest of the research community in music detection. This challenge was motivated by the need for speech and/or music detection for real-world tasks such as audio broadcasts where music can co-exists with speech and noise or may occur alone. Detecting audio segments with music can help apply music-specific methods on audio signals to achieve the desired impact~\cite{mirex_challenge}. The MIREX challenge has four tasks: Music Detection, Speech Detection, Music~\& Speech Detection, and Music Relative Loudness Estimation~\cite{mirex_challenge}. A CNN model with Mel-scale kernel was leveraged for music detection in broadcasting audio~\cite{jang2019music}. The Mel-scale kernel in CNN layers helped in learning robust features where the Mel-scale dictates the kernel size. This model was trained on 52 hours of mixed broadcast data containing approximately 50\% music; 24 hours of broadcast audio with a music ratio 50-76\% was used as a test set. This test set included representative genres for broadcast audio in English, Spanish, and Korean languages~\cite{jang2019music}.

A recent study~\cite{de2019exploring} conducted a comparative assessment of different convolutional, recurrent, and hybrid deep neural network architecture for music detection trained on mel-spectral features on Audio Set balanced training set~\cite{gemmeke2017audio}. Further, music detection using one neural network and two neural networks were compared. In the second case, one neural network was trained to detect speech and another one for detecting music~\cite{de2019exploring}. The study included fully connected, convolutional, and long short-term memory (LSTM) networks where the hybrid convolutional-LSTM model lead to the best performance~\cite{de2019exploring}. One of the systems~\cite{jang2018music} submitted at the MIREX 2018~\cite{mirex_challenge} leveraged a CNN with Mel-scale kernels for music detection and a bidirectional GRU model trained on Mel-spectrograms for speech detection. Another CNN system~\cite{melendez2018music} was submitted at MIREX 2018 for detecting music and estimating the music relative loudness. As music relative loudness represents the loudness corresponding to music content at each frame, thresholding the estimated loudness resulted in music/no-music decision~\cite{melendez2018music}.

A novel approach was recently developed for synthesizing the training datasets for a Convolutional Recurrent Neural Network (CRNN) model for speech and music detection which outperformed state-of-the-art approaches~\cite{venkatesh2021artificially}. This work provided an effective method for synthesizing a training dataset for speech/music classification. Real-time speech/music classification was proposed using line spectral frequencies and zero-crossing-based features extracted from audio frames~\cite{el2000speech}. Chroma features, which represent tonality in an audio signal, were applied for the task of speech/music discrimination~\cite{sell2014music}. Experiments show that chroma features outperform state-of-the-art features for speech/music classification on multiple corpora. Experiments on a broadcast news corpus validated the efficacy of chroma features for high-precision music detection even in mismatched acoustic conditions~\cite{sell2014music}.
\vspace{-4mm}
\section{ Proposed Approach}
\label{sec:proposed}
\vspace{-3mm}
\subsection{Training Dataset}
\label{sec:training_set}
\vspace{-2mm}
In this work, we leveraged Google's Audio Set dataset~\cite{gemmeke2017audio} to train MusicNet. This dataset contains 527 audio events including music. Each clip comes with labels containing names of audio events contained in it. The data is highly skewed in terms of speech and music classes. Both speech and music appear in almost a million clips while the rest of the audio events occur with 100 to 1000 clips for each class. Fortunately, roughly 50\% of the Audio Set clips had music as one of the labels which gives us enough data to train MusicNet. During training, audio clips with music as one of the labels were treated as music while the rest of the clips were treated as non-music. MusicNet was trained on a clip-level which means the loss is computed across all audio clips in a given training batch. We downsampled all audio to 16kHz as our application is targeted for wideband audio\footnote{\url{https://en.wikipedia.org/wiki/Wideband audio}}. MusicNet is trained to classify 9 second audio clips into music or no-music binary classification. Note that the 9 second window results in an average latency of ~4.5 seconds to act on detected music, which is sufficiently low for use in a real-time communication systems, e.g., the user can be prompted to switch to a speech enhancement mode that lets music pass through the noise suppression, or to use an audio codec that is optimized for music quality.  
\vspace{-3mm}
\subsection{Test set}
\label{sec:test set}
\vspace{-3mm}
We collected real-world test clips containing music with or without clean and/or background noise. We also added clips with only clean speech, noisy speech and/or background noises in the test set as no-music examples. We chose the blind test set containing real recordings with music in the background from the 3rd Deep Noise Suppression Challenge~\cite{ch2021interspeech}. We also included music clips from the publicly available Freesound dataset~\cite{fonseca2017freesound}. Our test set consisted of 1000 real test clips where each clip has approximately a 10 second duration. We added test clips from different instruments such as Piano, Guitar, Violin, Drums, Saxophone, Flute, Cello, Clarinet, Trumpet, Harp, etc. and different background music genres namely Rock, Pop, Country, R\&B and Soul, Hip Hop/Rap, Electronic, Jazz, Blues, Classical \& Opera, Folk, etc. We synthesized some scenarios for three popular instruments (from music lessons) Piano, Guitar, and Violin: (i) 20 clips for each instrument type containing only instrumental music. (ii) 10 test clips for each instrument type containing clean speech mixed with instrumental music. In addition, we had three signal to music ratios (SMRs) to account for three different gains for the musical instruments. In this way, we get 270 test clips (3 instruments * 30 clips * 3 SMRs). To this, we added 300 test clips without music where 150 test clips had only clean speech and 150 test clips had noisy speech. Noisy speech and clean speech clips were chosen from the 3rd Deep Noise Suppression Challenge as described in~\cite{ch2021interspeech}. Two expert listeners independently labeled the test set to ensure that the test set is strongly labeled for music/no-music classification. Our test set consists of music with clean speech, music with noisy speech, music with noise, only clean speech, only noisy speech, and only noise. The test set was collected through crowd sourcing using a variety of recording devices in both headset and speaker mode. It contains different musical instruments, background music, English speech, and non-English speech with or without background music. Our test set is designed to represent real-world scenarios for video/calls meetings.

Note that the annual event detection challenge DCASE (\url{https://dcase.community}) has N=1378 clips for their test set, but this is spread over 10 event types. Our test set of N=1000 clips for just 1 event type (music) gives a much higher number of clips per event type
%
%
\vspace{-3mm}
\subsection{State-of-the-art Models}
\vspace{-3mm}
We compared the performance of MusicNet with 20 state-of-the-art models described in~\cite{kong2020panns}. To reduce the experimental efforts and make our results reproducible, we leveraged the pretrained models made available on Github~\cite{github_kong2020panns}. These state-of-the-art models referred to as {PANNs} are a set of convolutional neural networks (CNNs) with different complexities taking raw audio waveform or a variety of spectral features as input and producing probabilities for each of the 527 acoustic classes contained in the Audio Set balanced training set. These PANNs models were trained using a balanced set of Audio Set training data; we did no further training on the models. 
In total, all PANN models were evaluated on 6 acoustic tasks where they showed good performance, thus validating their robustness for learning inherent structure in audio signals. Furthermore, performance and computational complexity trade-offs are studied for these models~\cite{kong2020panns}.
\begin{table}[!t!]
  \begin{center}
    \caption{MusicNet architecture}
    \label{tab:table1}
    \begin{tabular}{|l|r|}
    \hline
      \textbf{Layer} & \textbf{Output dimension} \\
      \hline

\hline
Featurization Conv layer  & 144 000 x 1\\
      \hline
      Conv: 32, (3 x 3), `ReLU' & 900 x 120 x 32 \\
      MaxPool: (2 x 2), Dropout(0.3) & 450 x 60 x 32 \\
      \hline
      Conv: 32, (3 x 3), `ReLU' & 450 x 60 x 32 \\
      MaxPool: (2 x 2), Dropout(0.3) & 225 x 30 x 32 \\
      \hline
      Conv: 32, (3 x 3), `ReLU' & 225 x 30 x 32 \\
      MaxPool: (2 x 2), Dropout(0.3) & 112 x 15 x 32 \\
      \hline
      Conv: 64, (3 x 3), `ReLU' & 112 x 15 x 64 \\
      GlobalMaxPool & 1 x 64 \\
      \hline
      Dense: 64, `ReLU' & 1 x 64 \\
      Dense: 64, `ReLU' & 1 x 64 \\
      Dense: 1  & 1 x 2 \\
      \hline
    \end{tabular}
  \end{center}
\end{table}
\vspace{-3mm}
\subsection{Proposed Model}
\label{ssec:models}
\vspace{-3mm}

We built a binary music classifier for detecting music in 9 second clips. To this end, we explored
different configurations of the CNN-based models. The MusicNet architecture for our best performing
model is shown in Table~\ref{tab:table1}. The input to the model is a 9 second raw audio waveform.
The first layer of the model is the featurization layer which converts the raw waveform to real and imaginary
parts of the spectrum, computes the absolute magnitude of the spectrum, multiplies it with the Mel
weights to get the Mel-spectrum, and finally takes the log to get the LogMel spectrum. The audio frames from the 9 second clips are stacked so that we have 900 x 320 dimensions for raw waveform (considering the overlap of 160 samples) input,
which is fed to two 1-D convolutional layers, one each for the real and imaginary part of the spectrum.

In the featurization layer, we learn two weight matrices to convert the raw waveform to real and
imaginary parts of the spectrum and keep the Mel-weights fixed. We used the librosa Mel-weights in our model. We take log power Mel-spectrogram with 120 Mel bands computed over a 9 second clip sampled at 16 kHz with a frame size
of 20 ms and a hop length of 10 ms. This results in an input dimension of 900 x 120. We train the
model using a batch size of 32 with the Adam optimizer and binary cross-entropy loss function until
the loss is saturated.

We experimented by adding batch normalization layers after every Conv layer in
Table~\ref{tab:table1}. However, adding batch normalization worsens the overall classification
accuracy and adversely affected the detection of some musical instruments. We explored other network
architectures among which MusicNet as shown in Table~\ref{tab:table1} generalized the best with
the least computational complexity.

\vspace{-3mm}
\subsection{Real-time Communication Pipeline in Production}
\label{ssec:production}
\vspace{-3mm}

We deploy the trained model to a wide variety of edge devices, ranging from Windows, Mac, and Linux PCs to
mobile phones (Android/iOS). For portability, we convert our model into the ONNX format so it can
leverage the ONNX Runtime C++ library for client-side inference. Fast real-time inference and low
CPU usage are critical requirements for production, and the model size is important when deploying
MusicNet to mobile devices over unreliable and/or metered networks. MusicNet is a small model
with fast inference and accurate classification and hence is well-suited for our application.

\vspace{-5mm}
\subsubsection{Featurization}
\vspace{-3mm}

MusicNet takes 16 kHz single-channel raw audio waveform and computes the 120-band
Mel spectrogram for each 20 ms frame with a hop size of 10 ms. While inference runtimes like
TensorFlow Lite and ONNX Runtime help with the model portability, porting the featurization pipeline,
in general, remains a challenge. In particular, STFT implementations in the popular Python libraries
like numpy, librosa, and TensorFlow all have different defaults for the parameters like windowing
and padding, or require support for complex numbers. This leads to code portability issues and often
leads to numerical mismatches between the Python featurization in training and its C++ implementation
in the product. In addition, integration and maintenance of the featurization code creates barriers for iterative development of new production models. To facilitate easier
integration, maintenance and updates to the production model, we build the featurization as part of
the model. We implemented featurization as 1-D convolutional layer in PyTorch. We explained the data
flow in the featurization layer in Section~\ref{ssec:models}. The in-model featurization enables a
portable ONNX model which takes raw 9 second audio and outputs the probability of music.

%
\vspace{-3mm}
\section{Results~\& Discussions}
\label{sec:results}
\vspace{-3mm}

We compare MusicNet to 20 PANN models from~\cite{kong2020panns}. To make the comparison
fair, we convert PANN models using the code from~\cite{github_kong2020panns} to the ONNX format and
benchmark them on ONNX Runtime v1.1, which is the version we use on client devices in
production. Our use case requires a high TPR at a low FPR=0.1\%. However, the overall AUC score is of
less importance as our operating point is decided based on the target application and how many false positives per hour are acceptable. 
Classifiers like Cnn14\_DecisionLevelMax have an AUC=0.99 but TPR=7.3\% at FPR=0.1\%, which is unacceptable performance and a good example why using AUC to select music classifiers is insufficient.
To ensure the best
customer experience, a FPR=0.1\% is chosen which translates to one false detection of the music events
every 3 hours. As Table \ref{tab:onnx_all} shows, at the chosen operating point of FPR=0.1\% MusicNet achieves the best performance of TPR=81.3\%.

Table~\ref{tab:onnx_all} also shows the inference time for 9-second audio clips measured on Intel
i7-1065G7 CPU running Windows 10 and ONNX Runtime 1.1. MusicNet is faster than the best
PANN model, and also is the smallest with a size of 0.22MB for an uncompressed ONNX file.

The key reasons why MusicNet performs better than other methods are: (1) We use 9 seconds of raw audio to extract 120 bands Mel spectrogram. Other methods use shorter audio lengths with fewer Mel bands that work well for speech, but not for detecting music. (2) We avoid any kind of batch normalization in our network as it makes the network insensitive to level variations in the embedding space, which is important to detect music. It also helps the model generalize better on challenging and realistic test set. (3) We took reasonable steps to ensure the generalizability of the model by using the right amount of dropout, filter lengths and the number of filters. MusicNet is much smaller than MobileNet V1 and V2 in terms of the memory and yet achieves better accuracy.
\begin{table}[t!]
  \begin{center}
 \vspace{-3mm}
 \caption{Inference time per 10-second audio frame, ONNX file size, and accuracy of the models in the study.}
 \vspace{-3mm}
 \label{tab:onnx_all}
 \begin{tabular}{|l|r|r|r|r|}
   \hline
   \textbf{Model} & \hspace{-1mm}\textbf{Inference} & \textbf{Size,} & \textbf{TPR} & \textbf{AUC} \\
        \textbf{} & \textbf{time, ms} & \textbf{MB} & \hspace{-1mm}\textbf{at 0.1\%} & \textbf{} \\
        \textbf{} & \textbf{} & \textbf{} & \textbf{FPR} & \textbf{} \\
   \hline \textbf{Proposed Model} & \textbf{11.1} & \textbf{0.2} & \textbf{81.3\%} & 0.97 \\
   \hline         Wavegram\_Cnn14 &  285.2 &  308.9 & 61.8\% & 0.95 \\
   \hline                 Cnn14\_ &        &        &        &      \\
    DecisionLevelMax\hspace{-1mm} &  249.4 &  308.1 &  7.3\% & 0.99 \\
   \hline           Cnn14\_emb512 &  242.4 &  293.0 & 67.6\% & 0.98 \\
   \hline           Cnn14\_emb128 &  239.6 &  289.2 & 40.9\% & 0.99 \\
   \hline            Cnn14\_emb32 &  240.8 &  288.3 & 57.9\% & 0.99 \\
   \hline                   Cnn14 &  240.1 &  308.1 & 51.7\% & 0.99 \\
   \hline                   Cnn10 &  163.2 &   19.9 & 48.9\% & 0.99 \\
   \hline                    Cnn6 &  140.7 &   18.5 & 80.7\% & 0.98 \\
   \hline              Res1dNet51 &  480.0 &  406.7 & 23.8\% & 0.94 \\
   \hline              Res1dNet31 &  271.8 &  307.2 & 24.9\% & 0.93 \\
   \hline                ResNet54 &  340.8 &  398.2 & 75.3\% & 0.99 \\
   \hline                ResNet38 &  282.3 &  281.6 & 46.4\% & 0.98 \\
   \hline                ResNet22 &  171.2 &  243.0 & 62.9\% & 0.98 \\
   \hline                DaiNet19 &  238.7 &   16.8 & 45.7\% & 0.95 \\
   \hline                LeeNet24 &  212.2 &   38.2 & 28.1\% & 0.93 \\
   \hline                LeeNet11 &   44.1 &    2.9 & 25.5\% & 0.95 \\
   \hline             MobileNetV2 &   14.4 &   15.7 & 46.1\% & 0.96 \\
   \hline             MobileNetV1 &   15.1 &   18.4 & 52.4\% & 0.98 \\
   \hline
 \end{tabular}
  \end{center}
\end{table}

\vspace{-3mm}
\section{Conclusions}
\vspace{-3mm}

We present a model for real-time background music detection and compare it with 20 other
state-of-the-art models. MusicNet is 10x smaller than the next smallest model in the
study, and is 35\% faster than the fastest PANN classifier. MusicNet achieves the best
TPR=81.3\% at a FPR=0.1\% operating point. Our model is portable
to a wide range of client devices and incorporates Mel-spectrogram featurization as part of the
model definition.

To validate the model performance in our use case, we created a representative test set of music
clips from different instruments and genres with and without speech and background noises.

Future work can include optimizing the MusicNet model to achieve TPR $\ge$ 95\% at FPR=0.1\% while further
reducing the inference time.

%
%
\bibliographystyle{IEEEbib}
\bibliography{refs}

\end{document}